\documentclass[superscriptaddress,
reprint,longbibliography,
preprintnumbers,nofootinbib,
amsmath,amssymb,aps,prb]{revtex4-1}

\pdfoutput=1

\usepackage[colorlinks = true,
            linkcolor = blue,
            urlcolor  = blue,
            citecolor =red,
            anchorcolor = blue]{hyperref}
            
\usepackage{latexsym}
\usepackage{lineno}
\usepackage{comment}
\usepackage{microtype}
\usepackage{url}
\usepackage{graphicx}
\usepackage{verbatim}
\usepackage{multirow}
\usepackage{amsmath}
\newcommand{\beq}{\begin{eqnarray}}
\newcommand{\eeq}{\end{eqnarray}}
\usepackage{mathrsfs}
\usepackage{float}
\usepackage[usenames, dvipsnames]{color}

\usepackage{mathtools}
\usepackage{slashed}
\usepackage{physics}	
\usepackage{graphicx}   
\usepackage{epstopdf}
\usepackage{subfigure}  
\usepackage{hyperref}   
\usepackage{bbold}
\usepackage{wasysym}
\usepackage{amssymb}
\usepackage{feynmp}
\usepackage[symbol]{footmisc}

\usepackage{tikz}
\usetikzlibrary{decorations.pathmorphing}
\usetikzlibrary{shapes.misc}
\tikzset{cross/.style={cross out, draw=black, minimum size=8*(#1-\pgflinewidth), inner sep=0pt, outer sep=0pt},
cross/.default={1pt}}
\usetikzlibrary{patterns,math}
\begin{document}

\title{\Large Universal $L^{-3}$ finite-size effects in the viscoelasticity \\ of confined liquids and amorphous media}

\author{Anthony E. Phillips}
\thanks{a.e.phillips@qmul.ac.uk}
\affiliation{School of Physics and Astronomy, Queen Mary University of London, U.K.}
\author{Matteo Baggioli}
\thanks{matteo.baggioli@uam.es}
\affiliation{Wilczek Quantum Center, School of Physics and Astronomy, Shanghai Jiao Tong University, Shanghai 200240, China.}
\affiliation{Shanghai Research Center for Quantum Sciences, Shanghai 201315.}
\affiliation{Instituto de Fisica Teorica UAM/CSIC, c/Nicolas Cabrera 13-15,
Universidad Autonoma de Madrid, Cantoblanco, 28049 Madrid, Spain.}
\author{Timothy W. Sirk}
\thanks{timothy.w.sirk.civ@mail.mil}
\affiliation{Polymers Branch, US Army Research Laboratory, Aberdeen Proving Ground, MD 21005, USA}
\author{Kostya Trachenko}
\thanks{k.trachenko@qmul.ac.uk}
\affiliation{School of Physics and Astronomy, Queen Mary University of London, U.K.}
\author{Alessio Zaccone}
\thanks{alessio.zaccone@unimi.it}
\affiliation{Department of Physics ``A. Pontremoli", University of Milan, via Celoria 16, 20133 Milan, Italy.}
\affiliation{Cavendish Laboratory, University of Cambridge, JJ Thomson Avenue, CB30HE Cambridge, U.K.}

\begin{abstract}
We present a theory of viscoelasticity of amorphous media, which takes into account the effects of confinement along one of three spatial dimensions.
The framework is based on the nonaffine extension of lattice dynamics to amorphous systems, or nonaffine response theory. The size effects due to the confinement are taken into account via the nonaffine part of the shear storage modulus $G'$. The nonaffine contribution is written as a sum over modes in $k$-space. With a rigorous argument based on the analysis of the $k$-space integral over modes, it is shown that the confinement size $L$ in one spatial dimension, e.g. the $z$ axis, leads to a infrared cut-off for the modes contributing to the nonaffine (softening) correction to the modulus that scales as $L^{-3}$. Corrections for finite sample size $D$ in the two perpendicular dimensions scale as $\sim (L/D)^4$, and are negligible for $L \ll D$. For liquids it is predicted that $G'\sim L^{-3}$ in agreement with a previous more approximate analysis, whereas for amorphous materials $G' \sim G'_{bulk} + \beta L^{-3}$. For the case of liquids, four different experimental systems are shown to be very well described by the $L^{-3}$ law, which also can explain previous simulation data of confined jammed granular packings.
\end{abstract}
\maketitle

\section{Introduction} Lattice dynamics can be extended to deal with disordered systems where the positions of atoms or molecules are completely random, to arrive at theoretical expressions for the elastic constants and for the viscoleastic moduli\cite{Lemaitre,Zaccone2011,Zaccone2013,Palyulin}. 
The resulting theoretical framework is sometimes referred to as nonaffine lattice dynamics or NALD\cite{Lemaitre,Zaccone2013}. The theory has proved effective in quantitatively describing elastic, viscoelastic and plastic response of systems as diverse as jammed random packings and random networks\cite{Zaccone2011}, glassy polymers\cite{Palyulin,Ness,Elder}, metallic glass\cite{PRB2014}, colloidal glasses\cite{Laurati2017,Denisov}, and perfect non-centrosymmetric crystals like quartz~\cite{Rodney}. Furthermore, NALD intrinsically takes into account long-range correlation phenomena\cite{Sheldon,Kob} that are present also in liquids and give rise to acoustic wave propagation.
Because of its microscopic character, and to its ability to represent contributions to elasticity in terms of eigenmodes of the Hessian or dynamical matrix of the systems, NALD is thus a promising framework to describe size-dependent effects due to confinement. 
Understanding these effects at the microscopic level is important for a wide variety of systems in condensed matter and materials physics~\cite{Alba,Petersen}, polymers~\cite{Napolitano1,Napolitano2,Riggleman2,Riggleman}, and amorphous and glassy systems~\cite{Cerveny,Smarajit}.

In this paper, we present a detailed analysis of size-dependent effects on the viscoelastic shear modulus of amorphous systems confined in one spatial dimension, including liquids and glasses. We evaluate the nonaffine integral over $k$-space generally by allowing the ``infrared'' limit to vary with polar angle $\theta$ and hence evaluate the scaling properties of the nonaffine contribution to the shear storage modulus. 
An analysis of four published experimental data sets shows that this scaling law is shared by many different systems, and remains valid for arbitrary chemical composition and microscopic or mesoscopic structure of the system.
Although we focus on linear viscoelasticity, these results could be useful also for understanding of plasticity of confined systems~\cite{Weygand}, and also for understanding mechanical fragmentation processes in dispersed, colloidal and biological systems, where mesoscopic aggregates display size-dependent mechanical properties~\cite{Conchuir}.

\section{Nonaffine viscoelastic theory}
The usual starting point is the equation of motion of a microscopic building block, i.e. an atom or a molecule for atomic liquids or molecular liquids, respectively. In the case of polymers, the building block could be identified with a monomer of the polymer chain\cite{Palyulin}. 
Following previous literature~\cite{Lemaitre,Zaccone2011}, we introduce the Hessian matrix of the system $\underline{\underline{H}}_{ij}=-\partial^2\mathcal{U}/\partial\underline{\mathring{q}}_i\partial\underline{\mathring{q}}_j$ and the affine force field $\underline{\Xi}_{i,\kappa\chi}=\partial\underline{f}_i/\partial\eta_{\kappa\chi}$, where $\eta_{\kappa\chi}$ is the strain tensor. For example, for simple shear deformation the $xy$ entry of tensor $\eta_{\kappa\chi}$ is given by a scalar $\gamma$, which coincides with the angle of deformation. 

As shown in previous works \cite{Lemaitre,Palyulin}, the equation of motion of an atom $i$ in a disordered medium subjected to an external strain, in mass-rescaled coordinates, can be written as:
\begin{equation}
\frac{d^2\underline{x}_i}{dt^2}+\nu\frac{d\underline{x}_i}{dt}+\underline{\underline{H}}_{ij}\underline{x}_j
=\underline{\Xi}_{i,\kappa\chi}\eta_{\kappa\chi} \label{eq1}
\end{equation}
where $\underline{\underline{\eta}}$ is the (Green-Saint Venant) strain tensor and $\nu$ is a microscopic friction coefficient which arises from dynamical couplings mediated by the anharmonicity of the pair potential. The term on the r.h.s. physically represents the effect of the disordered (non-centrosymmetric) environment leading to nonaffine motions: a net force acts on the atom $i$ in the affine position (i.e. the position prescribed by the external strain tensor $\eta_{\kappa\chi}$). 

In a disordered or non-centrosymmetric bonding environment, in order to keep mechanical equilibrium on all atoms throughout the deformation, an additional \textit{nonaffine} displacement is required in order to relax the force $f_{i}$ acting in the affine position. This displacement brings each atom $i$ to a new (nonaffine) position. 

The equation of motion Eq. \eqref{eq1} can also be derived from first principles, from a model particle-bath Hamiltonian as shown in previous work~\cite{Palyulin}.
Using standard manipulations (Fourier transformation and eigenmode decomposition from time to eigenfrequency~\cite{Lemaitre}), and applying the definition of mechanical stress as derivative of the energy, one obtains the following expression for the
viscoelastic (complex) elastic constants\cite{Lemaitre,Palyulin}:
\begin{equation}
C_{\alpha\beta\kappa\chi}(\omega)=C_{\alpha\beta\kappa\chi}^{\textit{Born}}-
\frac{1}{V}\sum_n\frac{\hat{\Xi}_{n,\alpha\beta}\hat{\Xi}_{n,\kappa\chi}}{\omega_{p,n}^2-\omega^2+i\omega\nu} \label{nonaffine}
\end{equation}
where $C_{\alpha\beta\kappa\chi}^{\textit{Born}}$ is the Born or affine part of the elastic constant, which is what survives in the infinite-frequency limit. Here, $\omega$ represents the oscillation frequency of the external strain field, whereas $\omega_p$ denotes the internal eigenfrequency of the liquid (which results, e.g., from diagonalization of the Hessian matrix~\cite{Palyulin}). We use the notation $\omega_{p}$ to differentiate the eigenfrequency from the external oscillation frequency $\omega$.

An atomistic expression for $G_{\infty} \equiv C_{xyxy}^{\textit{Born}}$ is provided by the well known Zwanzig-Mountain (ZM) formula~\cite{Zwanzig}, in terms of the pair potential $V(r)$ and the radial distribution function $g(r)$.
The sum over $n$ in Eq.\eqref{nonaffine} runs over all $3N$ degrees of freedom (given by the atomic or molecular building blocks with central-force  interactions). Also, we recognize the typical form of a Green's function, with an imaginary part given by damping and poles $\omega_{p,n}$ that correspond to the eigenfrequencies of the excitations.

At this point, we consider the dynamics of elastic waves in liquids. The propagation of longitudinal acoustic waves in liquids is of course a well known fact, with firmly established  both experimental and theoretical evidence of longitudinal acoustic dispersion relations\cite{Hubbard,Takeno,Hansen}. For transverse or shear acoustic waves in liquids, instead, there is no propagation below a characteristic wavenumber. Indeed, there is an onset value of $k$, that we shall denote $k_{g}$, above which these modes can propagate in liquids.
This represents a gapped momentum state seen in a number of different systems, including liquids, supercritical fluids, plasma, Keldysh-Schwinger theory, relativistic hydrodynamics, holographic and other models such as the sine-Gordon model\cite{BAGGIOLI20201}.
The gap increases with temperature and the inverse of liquid relaxation time (see, e.g., Refs. 
\cite{yang,Trachenko_Ga}).

Following the analytical steps presented in Ref.~\onlinecite{PNAS2020}, we arrive at the following expression for the frequency-dependent storage modulus $G'$,
\begin{align}
G^{*}(\omega)=&G_{\infty}
-B \int_{k_{\text{min}}}^{k_{D}}\frac{\omega_{p,L}^{2}(k)}{\omega_{p,L}^2(k)-\omega^2+i\omega \,\nu}k^{2}dk\nonumber\\
& - B\int_{k_{\text{min}}}^{k_{D}}\frac{\omega_{p,T}^{2}(k)}{\omega_{p,T}^2(k)-\omega^2+i\omega\, \nu}k^{2}dk\,, \label{integrals}
\end{align}
where the first integral represents the nonaffine (negative or softening) contribution due to longitudinal (L) acoustic modes, while the second integral represents the nonaffine (also softening) contribution due to the transverse (T) acoustic modes. In the above expression, $k_{\text{min}}$ is an ``infrared'' cutoff, which is $k_{\text{min}}=0$ for a standard bulk material, which can be considered as large in all spatial dimensions ($L=\infty$). $B$ is an arbitrary prefactor.
For liquids, $k_{\text{min}}=\max \left(k_g,k_{\text{conf}}\right)$, for the transverse modes, with $k_g$ the onset wavenumber for transverse phonons in liquids (the $k$-gap), and $k_{\text{conf}}$ is the wavenumber set by the confinement length (see below).

Upon taking the real part of $G^{*}$, which gives the storage modulus $G'$, and focusing on low external oscillation frequencies
$\omega \ll \omega_{p}$, in both integrals numerator and denominator cancel out, so that both integrals reduce to the same expression, a volume in $k$-space. Therefore, as anticipated above, the final low-frequency result does not depend on the actual form of $\omega_{p,L}(k)$, nor of $\omega_{p,T}(k)$, although the latter, in liquids, due to the $k$-gap, plays an important role.
In the experiments where the size effect of confinement is seen \cite{Noirez2011,Noirez_ionic,Martinoty,Riedo,private}, $k_g\ll\frac{1}{L}$. 
This is because the typical speed of sound $c$ in these systems (e.g., short-chain unentangled polymers) is of the order of 1000 $\mathrm{m\,s}^{-1}$ and the (Maxwell) relaxation time  $\tau$ is in the range (0.001--0.01) s.\cite{Winter} Hence $c\tau$ is in the range (0.1--1) m and is much larger than $L$ (which is on the submillimeter scale), therefore $k_g=\frac{1}{c\tau}\ll 1/L$ is justified. Similarly, for ionic liquids, we find that $c \sim 1000$ $\mathrm{m\,s}^{-1}$, while $\eta=0.38$~Pa and $G'=10$~Pa (data from Ref. [32]) which  gives $\tau = 0.038$ s, and again $1/c\tau \ll 1/L$ is satisfied. 
For the case of nanoconfined water, we lack data on the speed of sound, however, in that case $L=10^{-9}$ m. This would require an astronomically high speed of sound for the above condition to be violated, which is extremely unlikely.
Future work will address the situations where, instead, the above condition breaks down and $k_{min}$ lies inside the $k$-gap, which makes the Frenkel $k$-gap equation a directly relevant constraint in the above calculations.

Assuming that $k_{min}\approx \frac{1}{L}$, an approximation that we further expand on in the following section, we have
\begin{equation}
G'= G_{\infty} - \alpha \int_{1/L}^{k_{D}}k^{2}dk =G_{\infty} - \frac{\alpha}{3} k_{D}^{3} + \frac{\beta}{3} L^{-3}.   \label{result}
\end{equation}

For bulk (unconfined) liquids in thermodynamic equilibrium, it can be shown~\cite{Wittmer} that $G_{\infty} - \frac{\alpha}{3} k_{D}^{3}=0$, thus leaving:
\begin{equation}
G' = \beta' L^{-3}.
\end{equation}
For amorphous solids, instead, $G_{\infty} - \frac{\alpha}{3} k_{D}^{3} > 0$, and one has the final scaling on $L$ given by
\begin{equation}
  G' = G'_\text{bulk}+\beta' L^{-3},
  \label{result amorphous}
\end{equation}
where $G'_\text{bulk}$ is the value of shear modulus for unconfined, bulk samples. In equations (\ref{result}--\ref{result amorphous}), $\alpha$, $\beta$, and $\beta'$ are arbitrary prefactors.

Evidently, the scaling $L^{-3}$ is easier to observe in liquids, as in amorphous solids it may be overshadowed by noise. Nonetheless, it is important to present the theoretical prediction also for amorphous solids, as it may be verified experimentally or in simulations in future work.

\begin{figure}
    \centering
    \includegraphics
    {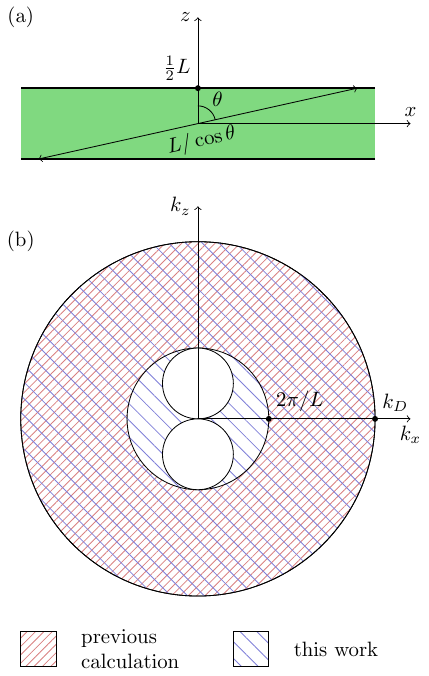}
    \caption{(a) Schematic section in real space of the confined cylindrical sample. (b) Geometry of the different regions over which the $k$-space integral (\ref{integrals}) can be taken; see full explanation in text. This is not to scale; in fact $k_D\gg 2\pi/L$. Both parts of this diagram have full rotational symmetry about the $z$ axis.} 
    \label{fig:geometry}
\end{figure}

\begin{figure}
    \centering
    \includegraphics[width=1.0 \linewidth]{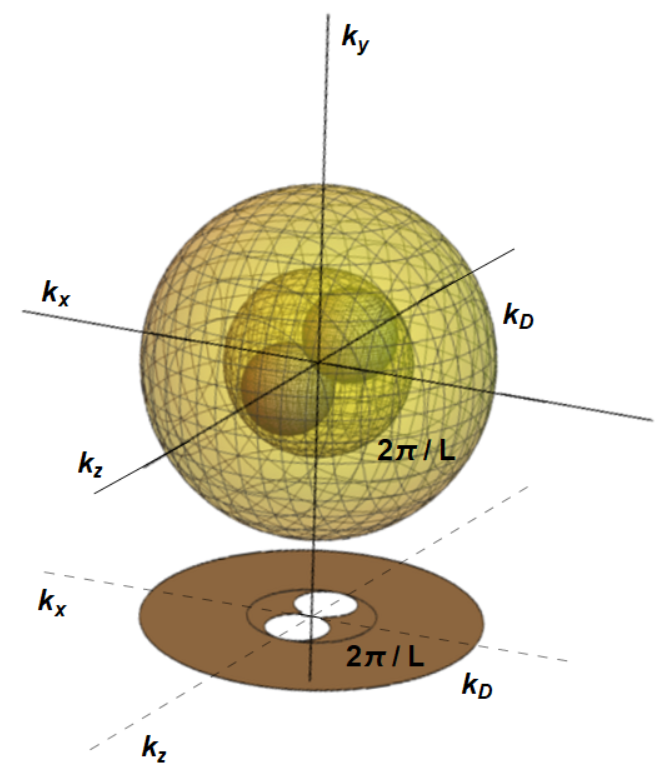}
    \caption{3D rendering of the geometry of integration in $k$-space for the confined system of Fig.~\ref{fig:geometry}.}
    \label{fig:sketch}
  \end{figure}

\section{General proof of the $L^{-3}$ law}

We consider a cylindrical system confined to length $L$ in the $z$ direction; for now we allow its extent in the perpendicular directions (that is, the cylinder's diameter) to be infinite. We use spherical polar coordinates, measuring the polar angle $\theta$ from the $z$ axis (Fig.~\ref{fig:geometry}(a)). Since our system has cylindrical symmetry, no quantities depend on the azimuthal angle $\phi$ and its origin is therefore arbitrary. In this notation the volume element in $k$-space is $\mathrm{d}V_k = k^2\mathrm{d}k\,\sin\theta\mathrm{d}\theta\,\mathrm{d}\phi$. If an integrand does not depend on $\theta$ or $\phi$, then $\mathrm{d}V_k = 4\pi k^2\mathrm{d}k$, demonstrating that the integrals in (\ref{integrals}) represent, to within a constant factor that can be absorbed into the prefactor, a volume in $k$-space.

If the system were unconfined, the lower limit on $k$ would simply be zero. Thus the region of allowable states would be a sphere in $k$-space, with radius equal to the Debye wavenumber $k_D$ and hence volume $\tfrac43\pi k_D^3$.\cite{Born,LandauFields,LandauStat,Kittel}
In our confined system, however, the maximum possible wavelength in the $z$ direction is approximately $\lambda_\text{max}\approx L$, giving a minimum wavevector of $k_\text{min}\approx 2\pi/L$. In our previous analysis,\cite{PNAS2020}, we made the simplifying approximation that the lower (``infrared'') limit of the $k$-space integral (\ref{integrals}) is $k_\text{min}$
\emph{regardless of the direction of propagation of the wave}. In this approximation, the lower limit is a spherical surface in $k$-space with radius $2\pi/L$, so that the integral should be taken over the pink narrow hatched volume in Fig.~\ref{fig:geometry}(b).

  Here we relax that assumption, showing that the $L^{-3}$ scaling holds even if we allow the lower limit to vary with $\theta$. If measured at an angle $\theta$ from the $z$ confinement axis, the extent of the confined medium is $L/\cos\theta$ (Fig.~\ref{fig:geometry}(a)). Taking this value, as before, to be the maximum allowed wavelength in that direction, we now have $k_\text{max} = 2\pi\cos\theta/L$. In the range $0\leq\theta\leq\pi$, this equation describes two spheres with radius $\pi/L$, centred at $(0,0,\pm\pi/L)$ in $k$-space. The integral (\ref{integrals}) must now be taken over the wide blue hatched volume in Fig.~\ref{fig:geometry}(b). A 3D rendered version of the same geometry is presented in Fig.~\ref{fig:sketch}.

  The volume of the two small spheres is
  \begin{equation}
    \label{eq:v_int}
    V_{k,\text{min}} = 2\times\tfrac43\pi\left(\frac{\pi}{L}\right)^3 = \frac{8\,\pi^4}{3\,L^3}.
  \end{equation}
  The allowable volume in $k$-space is therefore
  \begin{equation}
    \label{eq:v}
    V_k = \tfrac43\,\pi\, k_D^3 - \tfrac83\,\pi^4\,L^{-3}\,,
  \end{equation}
  displaying the same $L^{-3}$ scaling as derived previously.

  \begin{figure}
    \centering
    \includegraphics{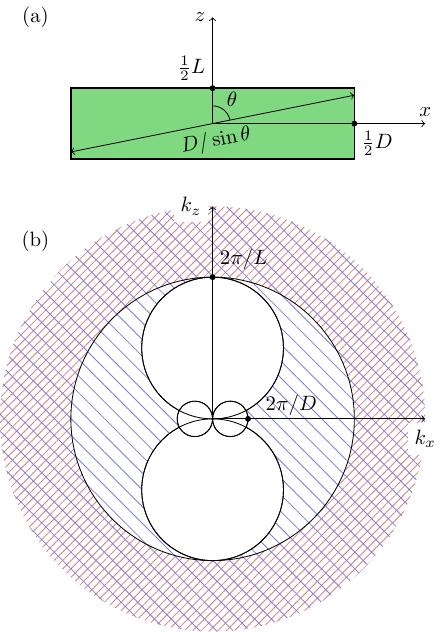}
    \caption{The same construction as Fig.~\ref{fig:geometry}, but allowing for the finite diameter $D$ of the cylinder. Here $D/L=4$. Compared to Fig.~\ref{fig:geometry}, (b) has been enlarged for clarity.}
    \label{fig:geometry2}
  \end{figure}
  
  We can extend this analysis still further by considering the effects of the finite cylinder diameter $D\gg L$ (Fig.~\ref{fig:geometry2}). The analysis proceeds as before except that the extent is now limited by $D$ rather than $L$ near $\theta=\pi/2$:
  \begin{equation}
    \label{eq:cylinder_extent}
    k_\text{min} =
    \begin{cases}
      2\pi\cos\theta/L & \lvert\tan\theta\rvert \leq D/L \\
      2\pi\sin\theta/D & \lvert\tan\theta\rvert \geq D/L
    \end{cases}.
  \end{equation}
  In this case, the infrared limit surface is the intersection of the two spheres and a toroidal shape. The internal volume is

  \begin{equation}
    \begin{split}
    \label{eq:v_int2}
    V_{k,\text{min}} &= 2\times2\pi\int_0^{\tan^{-1}(D/L)}\int_0^{2\pi\cos\theta/L}k^2\mathrm{d}k\,\sin\theta\mathrm{d}\theta \\
    &\qquad + 2\times2\pi\int_{\tan^{-1}(D/L)}^{\pi/2}\int_0^{2\pi\sin\theta/D}k^2\mathrm{d}k\,\sin\theta\mathrm{d}\theta \\
    &= \frac{8\pi^4}{3L^3}\bigg(1 - \frac{1 + L^2/D^2}{2(1 + D^2/L^2)^2} + \frac{2L^2/D^2}{1 + D^2/L^2}\\
      &\qquad + \frac{3L^2}{2D^2}\left[\frac{\pi}{2} - \tan^{-1}\left(\frac{D}{L}\right)\right]\bigg)\\
    &= \frac{8\pi^4}{3L^3}\Big(1 + 3(L/D)^4 + \mathcal{O}\big((L/D)^6\big)\Big).
  \end{split}
\end{equation}
As expected, this recovers the previous result \eqref{eq:v_int} in the limit as $D$ tends to infinity. Furthermore, it approaches this limit rather quickly, with the difference term being fourth-order in the aspect ratio $L/D$. In typical experiments, $L\approx 0.1D$, so that the difference from \eqref{eq:v_int} is negligible within experimental error.

We conclude that the $G' \sim L^{-3}$ scaling presented above in \eqref{result} is robust in two senses. First, it does not depend on the simplifying assumption previously made in Ref.~\onlinecite{PNAS2020}. Second, the correction term to allow for finite system size in the non-confined direction scales as the fourth power of the aspect ratio, making this correction negligible for typical experimental conditions where confinement is along the $z$ axis only.
 
\section{Discussion and comparison with experiments}
The above theory clarifies that the confinement between two plates is able to ``remove'' certain low-frequency normal mode collective oscillations of molecules, associated with the nonaffine motions (i.e. negative contributions to the elasticity), which are otherwise responsible for the fluid response of liquids under standard macroscopic (``unconfined'') conditions. These nonaffine motions are directly responsible for reducing the shear modulus, basically to zero in macroscopic liquids and to $G'_\text{bulk}$ in amorphous solids. Under confinement, instead, the shear modulus becomes non-zero for liquids, because these collective oscillations modes are suppressed, and the theory we (A.Z. and K.T.) have recently reported\cite{PNAS2020} provides the law by which the shear modulus grows upon reducing the confinement size $L$. In particular, the static shear modulus grows with the inverse cubic power of the confinement size $L$. 
For amorphous solids, the bulk shear modulus acquires an additional positive contribution $\sim L^{-3}$, due to confinement.

\begin{figure*}
	\centering
\includegraphics[width = 0.8 \linewidth]{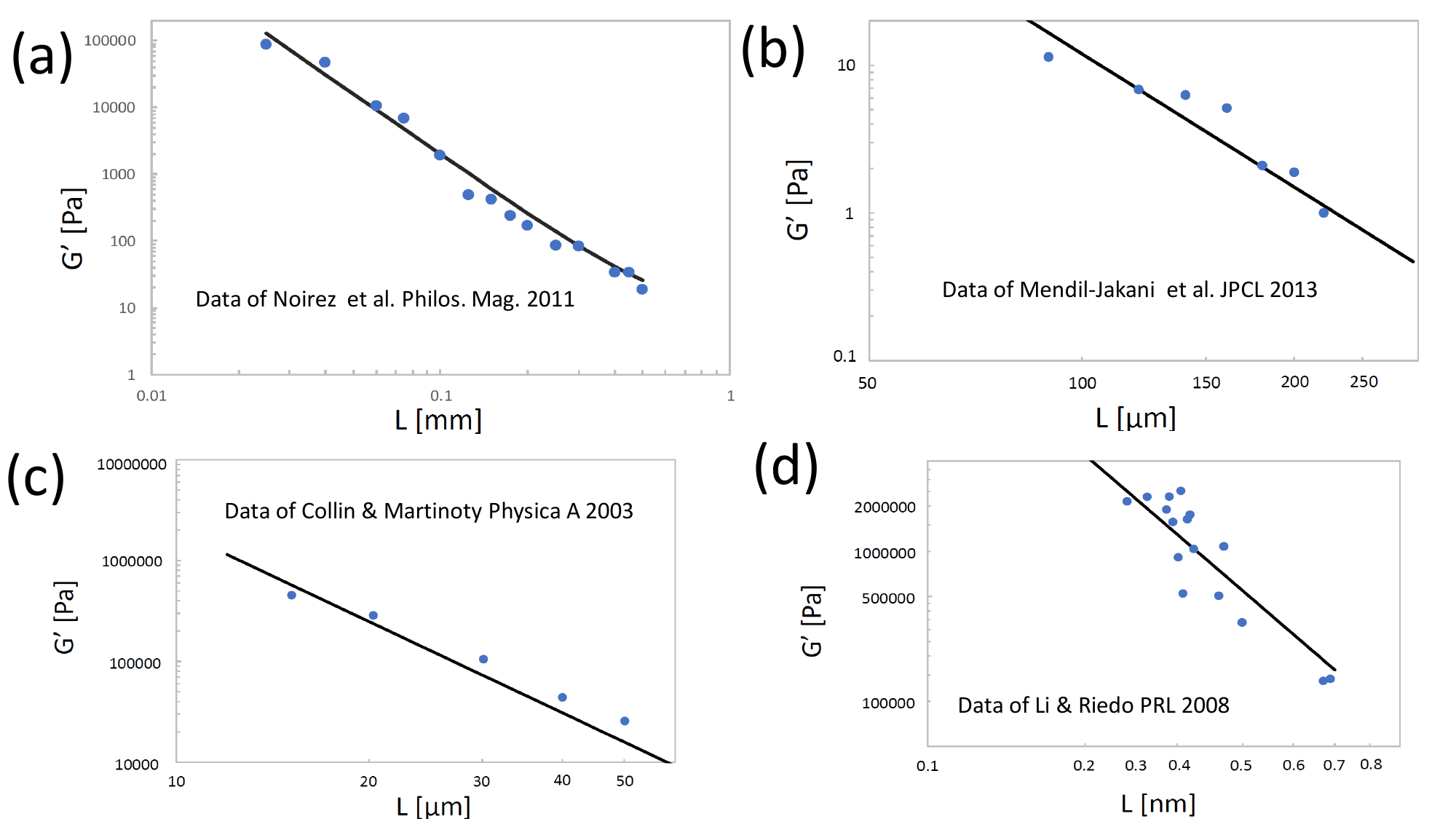}
	\caption{Experimental data of low-frequency shear modulus $G'$ versus confinement length $L$ for different systems: (a) short-chain (non-entangled) polybutylacrylate;\cite{Noirez2011} (b) an ionic liquid;\cite{Noirez_ionic} (c) short-chain (non-entangled) polystyrene melts;\cite{Martinoty} (d) nanoconfined water.\cite{Riedo} Circles represent experimental data while the solid line is the law $G' \sim L^{-3}$, with a prefactor determined by fitting to the data.}
	\label{fig:Nonaffine_Scheme}	
\end{figure*}

In Ref.~\onlinecite{PNAS2020}, the law $G' \sim L^{-3}$ was found to provide a good description of experimental data of the short-chain (non-entangled) polybutylacrylate upon varying the confinement $L$, using a conventional rheometer under good wetting conditions (see also Ref.~\onlinecite{Noirez2021}).
Given our conclusion in the previous section that this result should be robust across a wide range of media and experimental conditions, we now extend our comparison to more experimental systems.
In Fig.~\ref{fig:Nonaffine_Scheme}, we show the fits of this scaling law to three more experimental data sets. We observe that the scaling law agrees well
with experiments performed using short chain polymers, ionic liquids and nano-confined water.

The sources of the experimental data are given in the figure caption; we give here some further details of how these experimental studies relate to our own work.
In Ref.~\onlinecite{Martinoty}, experimental data on the plateau of $G'$ at different temperatures were also presented. The data in Fig. 2 of Ref.~\onlinecite{Martinoty} taken well above the glass transition at $T=114.30^{o}$C also follow the $L^{-3}$ scaling law reasonably well.
The experimental data on nanoconfined water were taken from Fig. 2(b') of Li and Riedo~\cite{Riedo}.  These data correspond to an oscillation  magnitude  $X_0$ of the AFM tip equal to $0.66$ nm, which is the middle value of those reported in Ref.~\onlinecite{Riedo} and presents the best compromise: the lower value $X_0=0.4$ nm is too close to the molecular size of water, whereas the larger value $1.32$ nm might be close to the nonlinear elastic regime and presents a less pronounced decay with $L$. 

Lastly, we note that prior work has addressed finite size scaling of the self-diffusion coefficient $D$ with hydrodynamic arguments, where a negative correction was predicted with a scaling proportional to $(\eta L)^{-1}$.\cite{dunweg1993,yeh2004} The results given here for $G'$ imply that a similar negative correction is necessary to recover the bulk value of dynamic viscosity $\eta$, due to the proportionality $G''/\omega\sim \eta'$ and the Kramers-Kronig relation between $G'$ and $G''$. Thus, the scaling of the viscosity described here, which was previously treated as a constant, shows the need for an additional higher order corrective term for the self-diffusion.

\section{Comparison with numerical simulations of
amorphous solids}
Over the past decades, many numerical simulations of the mechanical response of amorphous solids have been performed, in which the variation of the elastic moduli was studied for different ``coarse-graining'' sizes.\cite{Barrat2009,Mizuno2013,Rottler2019} In these systems, the shear modulus $G'$ is also found to increase upon decreasing the coarse-graining size. However, the overall effect is phenomenologically different from the finite-size effects discussed here. For example, in Ref.~\onlinecite{Rottler2019} $G'$ increases as $r_{c}^{-0.6}$, where $r_c$ is the coarse-graining size and hence has a very different exponent than the value $-3$ discussed here. The difference lies in the fact that the local sizes discussed in Ref.~\onlinecite{Rottler2019} do not refer to actual ``confinement'' as there are no physical boundaries involved, and the local regions are merely ``cropped'' within the same simulated material sample. In the absence of physical solid boundaries, the cut-off mechanism in $k$-space discussed here is not active, hence the $L^{-3}$ scaling in elasticity we discuss does not apply.

On the other hand, the situation presented by Goodrich, Liu and Nagel\cite{Goodrich} is very similar to the confinement effects we discuss above. In Ref.~\onlinecite{Goodrich}, jammed packings of frictionless soft spheres, one of the most widely studied models of amorphous solids,\cite{OHern} were studied by systematically varying the size of the simulation box that was subsequently subjected to shear deformation. 
In those simulations, it was found that the shear modulus increases by a positive correction that scales with $1/N$ upon decreasing the system's size $N$, where $N$ is the number of particles in the simulation box.
For Euclidean (non-fractal) systems in three dimensions, $N \sim L^{-3}$, hence the correction $\sim 1/N \sim L^{-3}$ exactly coincides with the correction that we predicted for amorphous solids, $G' \sim G'_{bulk} + \beta L^{-3}$. 
To our knowledge, this is the first theoretical derivation of the scaling $1/N$ for the shear modulus of confined jammed packings. This comparison more firmly establishes the ability of nonaffine response theory to predict the elastic properties of jammed systems. This theory has already provided a succcessful quantitative account of the shear modulus as a function of the distance to the jamming point, including prefactors.\cite{Zaccone2011}

\section{Conclusions}
In summary, we presented a microscopic theoretical framework for the size-dependent viscoelasticity of confined amorphous systems, both liquids and solids.
For the case of liquids, a previous approximate treatment~\cite{PNAS2020} has unveiled the surprising solid-like response under confinement, where the confinement effectively cuts off some nonaffine softening modes, leading to the scaling $G' \sim L^{-3}$ for the low-frequency shear modulus. 
In that earlier description, the integral over $k$-space, which provides the negative nonaffine correction, was evaluated approximately assuming 
that waves in any direction have the same maximal wavelength. 
Here, we presented a rigorous and general proof of the same result that takes the full $k$-space geometry of the problem into account, allowing the maximum wavelength to vary with the polar angle $\theta$. 
Our analysis shows that the $G' \sim L^{-3}$ law still holds when the initial approximation is relaxed. Furthermore, it is extremely robust with respect to finite sample size in the two perpendicular directions. These results are supported by an analysis of experimental data from the literature on four different liquids and complex fluids, all of which obey the $G' \sim L^{-3}$ law. 

We also derived a similar law for amorphous solids, with a predicted confinement-induced enhancement term in the low-frequency shear modulus that also scales with $L^{-3}$. This correction for amorphous solids is probably more challenging to verify, either experimentally or in simulations, but it may inspire further investigations.
On a more theoretical level, this result suggests that the limit of $G=0$, identically satisfied, for the zero-frequency shear modulus is attainable only in the thermodynamic limit of $L \rightarrow \infty$. This appears to broadly agree with earlier more formal results by Lebowitz~\cite{Lebowitz} and Ruelle~\cite{Ruelle}, recently re-discussed by Saw and Harrowell~\cite{Harrowell}, which point out that a nonzero shear modulus is the result of averaging over a constrained configuration space. 

Finally, it would be interesting in future work to study the interplay between confinement or boundary effects like those presented here and other low-$k$ phenomena in condensed matter such as hyperuniformity~\cite{Torquato} and its ramifications~\cite{Sciortino}.
Also, our theory predicts a $\sim L^{-3}$ positive correction term for the shear modulus of amorphous solids, which exactly agrees with the $1/N$ (where $N \sim L^{3}$ is the number of particles) finite-size correction term to the shear modulus observed numerically near the jamming transition of random jammed packings~\cite{Goodrich,OHern2,OHern1}.
Future work should be directed to further extending the above framework to deformation geometries other than shear, such as e.g. hydrostatic compression where nonaffine deformations can also be important for certain systems~\cite{Mizuno2013}.\\

\begin{acknowledgments}
M.B. acknowledges the support of the Shanghai Municipal Science and Technology Major Project (Grant No.2019SHZDZX01) and
of the Spanish MINECO ''Centro de Excelencia Severo Ochoa'' Programme under grant
SEV-2012-0249. C.S. is supported by the U.S. DOE grant number DE-FG02-05ER46236. A.Z. acknowledges financial support from US Army Research Laboratory and US Army Research Office through contract nr. W911NF-19-2-0055.
Discussions and input from Prof. Laurence Noirez are gratefully acknowledged.
\end{acknowledgments}
\bibliography{finite-size}
\onecolumngrid

\end{document}